\documentclass[aip,jcp,reprint,longbibliography]{revtex4-1} 
\usepackage{graphicx}
\usepackage{enumitem}
\usepackage{mathtools}
\usepackage{graphicx}   
\usepackage{color}
\usepackage[normalem]{ulem}
\usepackage{caption}
\usepackage{subcaption}
\usepackage{comment}
\usepackage{bm}
\usepackage{array}
\usepackage{booktabs}
\usepackage{multirow}

\usepackage{amsmath}
\usepackage{amssymb}

\usepackage{booktabs}

\usepackage[breaklinks=true]{hyperref}
\usepackage{breakcites}
\usepackage{algorithm}
\usepackage{algpseudocode}
\newcommand{\sm}{\textcolor{black}}
\newcommand{\pt}{\textcolor{black}}
\newcommand{\smm}{\textcolor{black}}

\begin{document}
\title{\sm{Thermodynamics-inspired Explanations of Artificial Intelligence}}

\author{Shams Mehdi}
 \affiliation{Biophysics Program and Institute for Physical Science and Technology,
 University of Maryland, College Park 20742, USA}
 
 \author{Pratyush Tiwary\footnote{Corresponding author.}} \email{ptiwary@umd.edu}
 \affiliation{Department of Chemistry and Biochemistry and Institute for Physical Science and Technology,
 University of Maryland, College Park 20742, USA.}
  \affiliation{University of Maryland Institute for Health Computing, Rockville, MD, USA}

	\date{\today}
	
	\begin{abstract}
\textbf{Abstract\newline }
In recent years, predictive machine learning methods have gained prominence in various scientific domains. However, due to their black-box nature, it is essential to establish trust in these models before accepting them as accurate. One promising strategy for assigning trust involves employing \smm{explanation} techniques that elucidate the rationale behind a black-box model's predictions in a manner that humans can understand. However, assessing the degree of human interpretability of the rationale generated by such methods is a nontrivial challenge. In this work, we introduce interpretation entropy as a universal solution for assessing the degree of human interpretability associated with any linear model. Using this concept and drawing inspiration from classical thermodynamics, we present \sm{Thermodynamics-inspired Explainable Representations of AI and other black-box Paradigms} (TERP), a method for generating accurate, and human-interpretable explanations for black-box predictions in a model-agnostic manner. To demonstrate the wide-ranging applicability of TERP, we successfully employ it to explain various black-box model architectures, including deep learning Autoencoders, Recurrent Neural Networks, and Convolutional Neural Networks, across diverse domains such as molecular simulations, text, and image classification.
\end{abstract}

	\maketitle

\section{Introduction}
\label{sec:Introduction}	

Performing predictions based on observed data is a general problem of interest in a wide range of scientific disciplines. Traditionally, scientists have tackled this problem by developing mathematical models that connect observations with predictions using their knowledge of the underlying physical processes. However, in many practical situations, constructing such explicit models is unfeasible due to a lack of system-specific information.\cite{dhar2013data} In recent years, an alternative class of purely data-driven approaches involving Artificial Intelligence (AI) has emerged with remarkable success.\cite{shalev2014understanding, lecun2015deep, davies2021advancing, carleo2019machine, mater2019deep, hamet2017artificial, baldi2001bioinformatics, brunton2022data} These methods are often referred to as black-box models, as they don't rely on a deep understanding of the system's inner workings and are designed to extract patterns directly from data. However, when it comes to making informed decisions and policies based on these models, this lack of understanding raises concerns.

\smm{Recently there has been significant progress in addressing this issue and the proposed approaches can be classified into two categories: } (a) AI models that are inherently explainable, or (b) post-hoc \smm{explanation} schemes for AI models that are not inherently explainable \smm{(XAI)}.\cite{rudin2019stop} Since most of the existing black-box AI are not inherently explainable, the latter class of methods has seen success in generating human comprehensible rationale behind AI predictions.\cite{molnar2020interpretable} \smm{Under this paradigm, different methods have been developed which can be black-box model-specific, or model-agnostic that generate global or locally valid explanations in the form of visual or feature importance attributions.{\cite{linardatos2020explainable,arrieta2020explainable,angelov2021explainable}}} For example, the LIME\cite{ribeiro2016should} method constructs a linear surrogate model that locally approximates the behavior of a black-box model. Coefficients associated with each feature of the constructed linear model are then used to attribute local feature importance. On the other hand, permutation-based \smm{explanation} methods such as SHAP\cite{lundberg2017unified} are designed to identify non-linear dependence among features at a higher computational cost. Other relevant \smm{explanation} approaches include gradient-based methods such as layer-wise relevance propagation (LRP),\cite{montavon2019layer} guided back-propagation,\cite{springenberg2014striving} integrated gradients\cite{sundararajan2017axiomatic} etc. Although these methods are designed to rationalize AI predictions, there is a potential issue regarding their level of human interpretability. For instance, when the rationalization involves a high number of correlated features, achieving high human interpretability and consequently establishing trust can be challenging. \smm{While quantifying the degree of human interpretability of AI explanation techniques has been an area of active research, the progress has been arguably limited. The progress so far includes methods that construct linear models to approximate AI models, by directly taking the number of model parameters or other similar approaches e.g, Akaike information criterion\cite{akaike1974new} or Bayesian information criterion\cite{schwarz1978estimating} as a proxy for human-interpretability.}

\begin{figure}
    \centering
    \includegraphics[width=\linewidth]{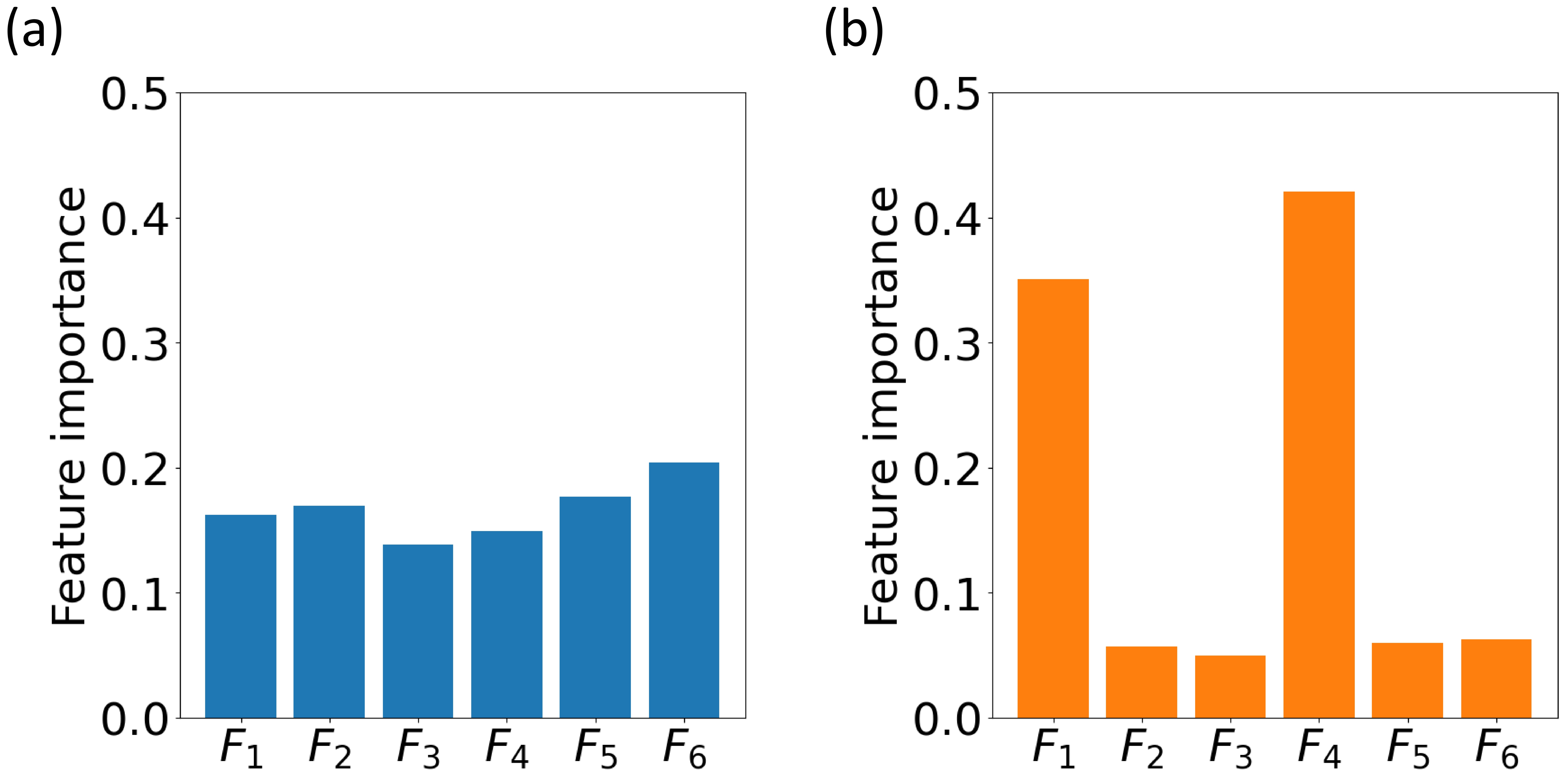}
    \caption{\textbf{Model complexity is not a good descriptor for human interpretability.} Illustrative linear models (a), \& (b) predict a target with the same level of accuracy. Both have the same number of model parameters (six), however, model (b) is significantly more human-interpretable than model (a). In model (b), two out of the six features stand out as most relevant for making predictions while it's difficult to identify relevant features for model (a).}
    \label{fig:int_illus} 
\end{figure}

One of the primary motivations behind our work is the recognition that model complexity can be an insufficient descriptor of human interpretability as shown in \textbf{Fig. } \ref{fig:int_illus}. In this case, if model complexity is used as a proxy for human interpretability, then both linear models shown in \textbf{Fig. } \ref{fig:int_illus}(a,b) will be assigned the same value as they both have the same number of model parameters. \smm{Indeed, previous studies \cite{miller1956magical, gigerenzer2009homo, chang2009reading} have revealed constraints in human cognition arising from a bottleneck in information processing capacity when subjected to different stimuli. Thus, we ground ourselves in the information-theoretic definition of entropy (\textbf{Sec. }\ref{sec:int_entropy}),\cite{bromiley2004shannon} and adopt a methodology that views linear model weights as a probability distribution. This allows us to assess differences in human interpretability among the different linear models by calculating a quantity similar to Shannon entropy. As illustrated in \textbf{Fig. }\ref{fig:int_illus}, it is evident that model (b) is significantly more understandable to humans compared to model (a). If both models exhibit equal accuracy, then a selection of model (b) over (a) is desirable, as it gives a few actionable strategies.} We solve this problem by introducing the concept of interpretation entropy for assessing the degree of human interpretability of any linear model. We show that under simple conditions, our definition of interpretation entropy addresses the shortcomings of complexity-based quantification.

Furthermore, we view the overall problem of AI model \smm{explanation} from the lens of classical thermodynamics.\cite{callen} It is known in thermodynamics that the equilibrium state of a system is characterized by a minimum in its Helmholtz Free Energy $F(T,V):=U-TS$. Here $U$, and $S$ represent the internal energy and entropy respectively of a system with a fixed number of particles $N$ at constant temperature $T$ and volume $V$. Similarly, we set up a formalism in this work where the optimality of an \smm{explanation} ($\zeta$) is assessed as a trade-off between its unfaithfulness ($\mathcal{U}$) to the underlying ground truth, and interpretation entropy ($\mathcal{S}$). Similar to $U$ and $S$ in classical thermodynamics, in our formalism $\mathcal{U}$ and $\mathcal{S}$ depend monotonically on each other. The strength of this trade-off can be tuned to identify the most stable \smm{explanation} using a parameter $\theta$, which plays a role similar to thermodynamic temperature $T$. For any choice of $\theta>0$, $\zeta$ is then guaranteed to have exactly one minimum characterized by a pair of values $\{\mathcal{U}, \mathcal{S}\}$ under certain conditions.

 We call our approach Thermodynamics-inspired Explainable Representations of AI and other black-box Paradigms (TERP). Owing to its model-agnostic implementation, TERP can be used for explaining predictions from any AI classifier. We demonstrate this generality by explaining the following black-box models in this work: (i) autoencoder based VAMPnets\cite{mardt2018vampnets} for tabular molecular data, (ii) self-attention based vision transformers for images\cite{dosovitskiy2020image} and, (iii) attention-based bidirectional long short-term memory (Att-BLSTM) for text\cite{zhou2016attention} classification. In particular, the first class of models belongs to an area of research undergoing rapid progress\cite{ma2005automatic,wang2020machine} involving molecular dynamics (MD) simulations.\cite{frenkel2001understanding, doerr2021torchmd, han2017deep, gao2020torchani, wang2021state,wang2020machine,beyerle2022quantifying, ribeiro2018reweighted,vanden2014transition,smith2020discovering, mehdi2022accelerating, beyerle2023recent,mehdi2024enhanced} 
As researchers with a keen interest in MD simulations, we have observed that the application of AI \smm{explanation} tools to AI models in this field has been very limited. Consequently, we believe that our proposed method TERP will prove valuable to the broader scientific community focused on this subject.

\section{Results}
\label{sec:Results}

\subsection{Interpretation Unfaithfulness ($\mathcal{U}$) for surrogate model construction}
\label{sec:unf}

Our starting point is some given dataset $\mathcal{X}$ and corresponding predictions $g$ coming from a black-box model. For a particular element $x\in \mathcal{X}$, we seek \smm{explanations} that are as human interpretable as possible while also being as faithful as possible to $g$ in the vicinity of $x$. \smm{We aim to address this problem of explaining $g$ by developing a linear approximation $F$ instead, which is more interpretable due to its linear construction. Specifically, we formulate F as a linear combination of an ordered set of representative features, $s=\{ s_1, s_2, ...,s_k,..., s_n \}$. Typically, these features are domain dependent} e.g, one-hot encoded superpixels for an image, keywords for text, and standardized values for tabular data. \sm{We demonstrate this in \textbf{Eq. } \ref{eq:simple model} below, where $F$ represents the linear approximation, $f_0$ is a constant, and $f_k$ comes from an ordered set of feature coefficients, $f=\{ f_1, f_2, ..., f_k,..., f_n \}$.}

\begin{equation} 
\begin{aligned} 
\label{eq:simple model}
\sm{F} = f_0 + \Sigma_{k=1}^n f_k s_k
\end{aligned} 
\end{equation}

Let's consider a specific problem where $x_0$ is a high-dimensional instance, and $g(x_0)$ is a black-box model prediction, for which an \smm{explanation} is needed. We first generate a neighborhood $\{ x_1, x_2, ... , x_N \}$ of $N$ samples by randomly perturbing the high-dimensional input space.\cite{molnar2020interpretable} A detailed discussion of neighborhood generation is provided in \textbf{Sec. \ref{sec:Methods}}. Afterwards, the black-box predictions $\{ g(x_1), g(x_2), ... , g(x_N) \}$ associated with each sample in the neighborhood are obtained. Subsequently, a linear, local surrogate model \sm{$F$} is constructed using linear regression with the loss function defined in \textbf{Eq. }\ref{eq:loss}.

\begin{multline}
\label{eq:loss}
    \mathcal{L} = \mathop{min}_{f_k} {\sum_{i=1}^N \Pi_i(x_0,x_i)\left[g(x_i)-(\sum_{k=1}^n f_k s_{ik})\right]^2}
\end{multline}

Here $\Pi_i(x_0, x_i) = e^{-d(x_0, x_i)^2/\sigma^2}$ is a Gaussian similarity measure, where $d$ is the distance between the explanation instance $x_0$ and a neighborhood sample $x_i$. In previous surrogate model construction approaches,\cite{ribeiro2016should} Euclidean distance in the continuous input feature space has been the typical choice for $d$. However, if the input space has several correlated or redundant features, a similarity measure based on Euclidean distance can be misleading.\cite{karagiannopoulos2004feature,liang1993regression} TERP addresses this problem by computing a one-dimensional (1-d) projection of the neighborhood using linear discriminant analysis\cite{izenman2008linear} (LDA), which removes redundancy and produces more accurate similarity. Such a projection encourages the formation of two clusters in a 1-d space, corresponding to in-class and not in-class data points respectively by minimizing within-class variance and maximizing between-class distances. Since the projected space is one-dimensional, there is no need to tune the hyperparameter, $\sigma$ in $\Pi_i(x_0, x_i) = e^{-d(x_0, x_i)^2/\sigma^2}$ as might be necessary in established methods and we can set $\sigma=1$. Advantages of LDA-based similarity have been highlighted for a practical problem in \textbf{Sec. }\ref{sec:AI augmented MD - VAMPnets} \textbf{Fig.} \ref{fig:vamp_01}(e-g).

Next, we introduce a meaningful unfaithfulness measure ($\mathcal{U}$) of the generated interpretation, computed from the correlation coefficient \textit{C} between linear model predictions (\sm{$F$}) and ground truth ($g$). For any interpretation, $C(\sm{F}, g) \in [-1,+1]$ and thus interpretation unfaithfulness is bounded, $\mathcal{U} \in [0,1]$

\begin{equation}
\label{eq:corrcoef}
   \mathcal{U} = 1 - |C(\sm{F}, g)|
\end{equation}

\sm{Using these definitions, we implement a forward feature selection scheme by first constructing $n$ linear models, each with $j=1$ non-zero coefficients. We use \textbf{Eq. } \ref{eq:corrcoef} to identify the feature responsible for the lowest $\mathcal{U}^{j=1}$. Here, the superscript $j=1$ highlights that $\mathcal{U}$ was calculated for a model with $j=1$ non-zero coefficients. We will follow this notation for other relevant quantities throughout this manuscript.} 

\sm{Afterwards, the selected feature is propagated to identify the best set of two features resulting in the lowest $\mathcal{U}^{j=2}$, and the scheme is continued until $\mathcal{U}^{j=n}$ is computed. Since a model with $j+1$ non-zero coefficients will be less or at best equally unfaithful as a model with $j$ non-zero coefficients defined in Eq. \ref{eq:simple model}, it can be observed that $\mathcal{U}$ monotonically decreases with $j$. The overall scheme generates $n$ distinct interpretations as $j$ goes from $1$ to $n$.}

\subsection{Interpretation Entropy ($\mathcal{S}$) for model selection}
\label{sec:int_entropy}

After identifying $n$ interpretations, our goal is to determine the optimal interpretation from this family of models. At this point, we introduce the definition of interpretation entropy $\mathcal{S}$ for quantifying the degree of human interpretability of any linear model. \sm{Given a linear model with an ordered set of feature coefficients $\{ f_1, f_2, ..., f_k,..., f_n \}$ among which $j$ are non-zero, we can define $\{ p_1, p_2, ..., p_k,..., p_n \}$, where $p_k := \frac{|f_k|}{\sum_{i=1}^{n}|f_i|}$}. The interpretation entropy is then defined as:

\begin{equation}
    \mathcal{S}\sm{^j} =  -\sum_{k=1}^{n} p_k \text{ log } p_k |\{ \text{log} p_k = 0 \text{ } \forall \text{ } p_k = 0 \}
\end{equation}

\pt{Here the subscript $j$ indicates that $\mathcal{S}$ was calculated for a model with $j$ non-zero coefficients.} It is easy to see that $p_k$ satisfies the properties of a probability distribution. Specifically, $p_k \geq 0$ and $\sum_{k=1}^{n}p_k = 1$.

\textbf{\textit{Lemma 01:}} Similar to the concept of \textit{self-information/surprisal} in information theory, the negative logarithm of $p_k$ from a fitted linear model can be defined as the self-interpretability penalty of that feature. Interpretation entropy is then computed as the expectation value of self-interpretability penalty of all the features as shown in \textbf{Eq. }\ref{eq:int_expec}. Using Jensen's inequality it is easy to show that $\mathcal{S}$ has an upper limit of $\text{log }k$ and we can normalize the definition so that $\mathcal{S}$ is bounded between $[0,1]$.

\begin{equation}
\label{eq:int_expec}
    \mathcal{S}\sm{^j} = \frac{-1}{\text{log }k} \sum_{k=1}^{n} p_k \text{ log } p_k = \frac{1}{\text{log }k}\mathbb{E}[-\text{log }p]
\end{equation}

This functional form of interpretability penalty \textit{i.e,} interpretation entropy ($\mathcal{S}$) encourages low values for a sharply peaked distribution of fitted weights indicating high human interpretability and vice-versa. Furthermore, if the features are independent, $\mathcal{S}$ has two very interesting properties expressed in the theorems below. The proofs of these two crucial theorems are provided in supporting information (SI).

\textbf{\textit{Theorem 01:}} $\mathcal{S}_k$ is a monotonically increasing function of the number of features ($k$).

\textbf{\textit{Theorem 02:}} $\mathcal{S}$ monotonically increases as $\mathcal{U}$ decreases.

\subsection{\sm{Free Energy ($\zeta$) for optimal explanation}}
\label{sec:optimality}

\begin{figure}
     \centering
         \includegraphics[width=\linewidth]{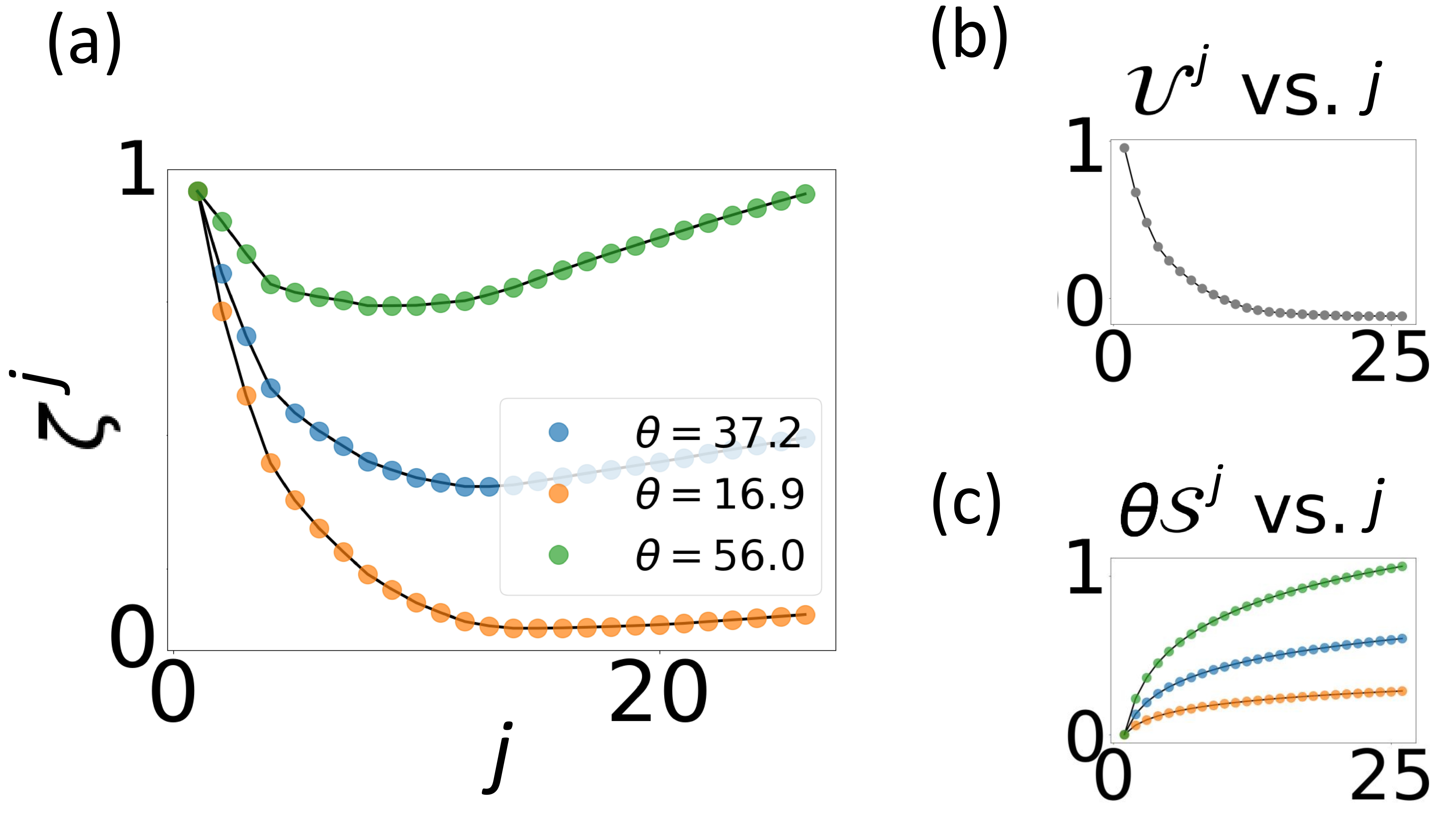}
         \caption{\textbf{Illustrative example highlighting properties of free energy $\zeta^j$, unfaithfulness $\mathcal{U}^j$, and interpretation entropy $\mathcal{S}^j$.} (a) Strength of $\mathcal{S}^j$ contribution to $\zeta^j$ can be tuned using $\theta$. $\zeta^j$ vs. $j$ plots for three different $\theta=37.2,16.9,56.0$ are shown resulting in a minimum at $j=12,14,8$ respectively. (b) $\mathcal{U}^j$ vs. $j$ remains unaffected by $\theta$. (c) $\mathcal{S}^j$ vs. $j$ plot shows that the strength of the trade-off can be tuned by $\theta$.}
         \label{fig:illustration}
\end{figure}

For an interpretation with $j$ non-zero coefficients, we now define free energy $\zeta\sm{^j}$ as a trade-off between $\mathcal{U}\sm{^j}$, and $\mathcal{S}\sm{^j}$ tunable by a parameter $\theta \geq 0$:

\begin{equation} 
\begin{aligned} 
\label{eq:zeta_functional}
\zeta\sm{^j}(f,\theta)=\mathcal{U}\sm{^j}+\theta \mathcal{S}\sm{^j}
\end{aligned} 
\end{equation}

\begin{figure*}[t!]
    \centering
    \includegraphics[width=\linewidth]{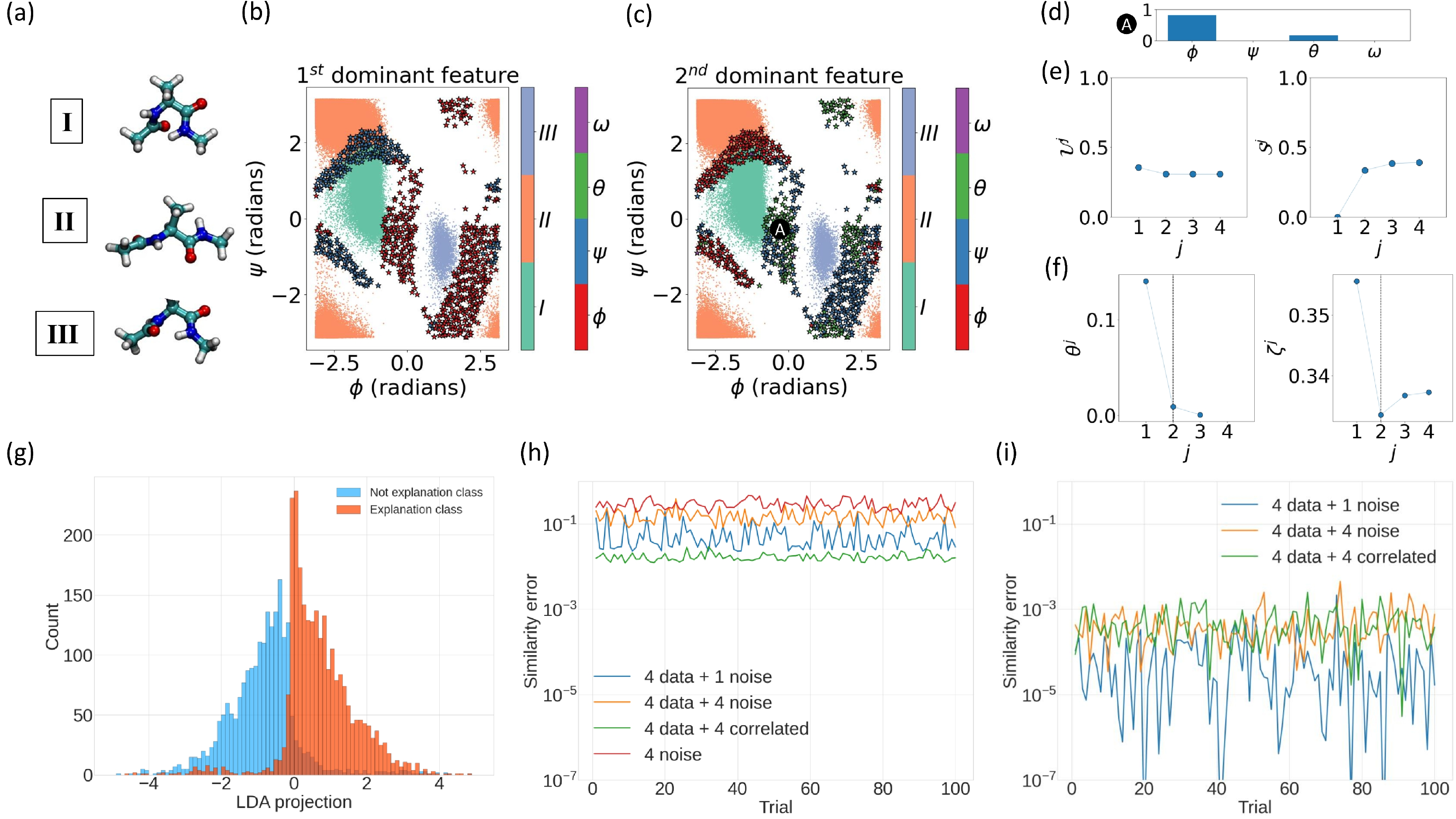}
    \caption{\textbf{Using TERP to explain VAMPnets for molecular dynamics simulations of alanine dipeptide in vacuum.} (a) Representative conformational states of alanine dipeptide labelled I, II, III. Projected converged states are highlighted in three different colors as obtained by VAMPnets along (b) ($\phi,\psi$) dihedral angles. \smm{$713$ different configurations are chosen for TERP and the first, and second dominant features are highlighted using colored ($\star$).} (d) Relative feature importance of a specific example point A. (e) High dimensional neighborhood data projected onto 1-d using LDA for improved similarity measure. \sm{Binarizing the class prediction probabilities of the neighborhood using a threshold of $0.5$ results in explanation and not explanation classes respectively. The LDA projection separates the two regimes of prediction probability, showing meaningful projection.} Average similarity error, $\Delta Pi$ (\textbf{Eq.} \ref{eq:sim}) per datapoint for (f) Euclidean, and (g) LDA based similarity respectively. Comparison between (f) and (g) shows minimal error for LDA based similarity, specifically demonstrated for an input space constructed from the four dihedral angles plus one pure noise, four pure noise, and four correlated features with partial noise respectively. The input space for no actual data and four pure noise features in (f) establishes a baseline, showing that the Euclidean similarity will include significant error even when one redundant feature is included. All the calculations were performed in 100 independent trials to appropriately examine the effects.}
    \label{fig:vamp_01} 
\end{figure*}

\sm{By writing an expression (\textbf{Eq.} \ref{eq:temp}) for the stationary value, $\Delta \zeta^j = \zeta^{j+1}-\zeta^j=0$, we can define characteristic temperatures $\theta^j$ at each $j \in [1,n-1]$. Essentially, $\theta^j= -\frac{\Delta \mathcal{U}^j}{\Delta \mathcal{S}^j}$ is a measure of change in unfaithfulness per unit change in interpretation entropy \pt{ for a model with $j$ non-zero coefficients}. 
This closely resembles the definition of thermodynamic temperature which is defined as the derivative of internal energy with respect to entropy. Afterwards, we identify the interpretation with $(j+1)$ non-zero coefficients that minimizes $(\theta^{j+1} - \theta^{j})=-(\frac{\Delta \mathcal{U}^{j+1}}{\Delta \mathcal{S}^{j+1}}-\frac{\Delta \mathcal{U}^{j}}{\Delta \mathcal{S}^{j}})$ as the optimal interpretation since it is guaranteed that $\zeta^{j+1}$ will preserve the lowest minimum among the set $\{\zeta^1, \zeta^2, ..., \zeta^j, ..., \zeta^n \}$ within the widest range of temperatures. Finally, we calculate optimal temperature, $\theta^o=\frac{\theta^{j+1}+\theta^{j}}{2}$ (any value within $\theta^j<\theta<\theta^{j+1}$ is equally valid since the optimal interpretation itself does not change) and generate the explanation as weights of this model. All $\zeta^j$ vs. $j$ plots shown in this manuscript are created using this definition of optimal temperature.}

\begin{equation}
\label{eq:temp}
\begin{split}
    \sm{\zeta^{j+1}-\zeta^j }&\sm{= (\mathcal{U}^{j+1}-\mathcal{U}^{j}) + \theta (\mathcal{S}^{j+1}-\mathcal{S}^j)}\\
    \sm{\Delta \zeta^j} &\sm{= \Delta \mathcal{U}^j + \theta \Delta \mathcal{S}^j}\\
    \sm{\theta^j} &\sm{= -\frac{\Delta \mathcal{U}^j}{\Delta \mathcal{S}^j}  \text{[By setting } \Delta \zeta^j=0\text{]}}\\
\end{split}
\end{equation}

Thus,
\begin{equation}
\label{eq:last}
\begin{split}
    \sm{\zeta^j}&\sm{= \mathcal{U}^{j} + (-\frac{\Delta \mathcal{U}^j}{\Delta \mathcal{S}^j}|_{\Delta \zeta^j = 0}) \mathcal{S}^j}\\
\end{split}
\end{equation}

This is again reminiscent of classical thermodynamics where a system's equilibrium configuration will in general vary with temperature but the coarse-grained metastable state description remains robust over a well-defined range of temperatures. In our framework, when $\theta=0$, $\zeta^j$ is minimized at $j=n$ interpretation or the model that maximizes unfaithfulness and completely ignores entropy. As $\theta$ is increased from zero, interpretation entropy contributes more to $\zeta^j$. Here, $(\theta^{j+1} - \theta^{j})$ is a measure of the stability of the $j$ non-zero coefficient interpretation. The complete TERP protocol is summarised in \textbf{Algorithm} \ref{alg1}.

\begin{algorithm}[H]
\caption{TERP protocol}
\begin{algorithmic}[1]
\State Generate neighborhood data by perturbing input features. Obtain associated black-box predictions.
\State Normalize, and then compute the similarity of the neighborhood samples using linear discriminant analysis (LDA).
\State Using ridge regression\cite{hoerl1970ridge} construct linear, surrogate models for all possible combinations of features at a specific $j$.
\State Implement forward feature selection by choosing the model with the lowest $\mathcal{U}^j$ at a specific $j$.
\State Compute $\mathcal{S}^j$ corresponding to all the chosen $j|\{0<j\leq n\}$ interpretations
\State Obtain the optimal \smm{explanation} by computing characteristic $\theta^j$ of the models and identifying minimum $(\theta^{j+1} - \theta^{j})$.
\end{algorithmic}
\label{alg1}
\end{algorithm}

\subsection{Application to AI-augmented MD: VAMPnets}
\label{sec:AI augmented MD - VAMPnets}

Variational approach for markov processes (VAMPnets) is a popular technique for analyzing molecular dynamics (MD) trajectories.\cite{mardt2018vampnets} VAMPnets can be used to featurize, transform inputs to a lower dimensional representation, and construct a markov state model\cite{bowman2013introduction} in an automated manner by maximizing the so called VAMP score. Detailed discussion of VAMPnets theory and parameters are provided in \textbf{Sec. \ref{sec:methods_vamp}}.

In this work, we trained a VAMPnets model on a standard toy system: alanine dipeptide in vacuum. An 8-dimensional input space with sines and cosines of all the dihedral angles $\phi, \psi, \theta, \omega$ was constructed and passed to VAMPnets. VAMPnets was able to identify three metastable states I, II, and III as shown in \textbf{Fig.} \ref{fig:vamp_01} (b,c).

To explain VAMPnets model predictions using TERP, \sm{we picked $713$ different configurations, some of which are near different transition states. To quantify data points as being a transition state, we use the criterion that the prediction probability for both classes should be higher than a threshold of $0.4$. From a physics perspective, the behavior of such molecular systems near the transition states is a very pertinent question.
Additionally, class prediction probability is the most sensitive at the transition state and if our method generates a meaningful local neighborhood it will include a broad distribution of probabilities resulting in highly accurate approximations to the black-box behavior. Thus, a correct analysis of the transition state ensemble will validate our similarity metric and overall neighborhood generation scheme.} 

\smm{We generated $5000$ neighborhood samples for each configuration and performed TERP by following \textbf{Algorithm} \ref{alg1}. In \textbf{Fig.} \ref{fig:vamp_01} (b,c) we highlight the first, and second most dominant features using colored stars ($\star$) identified by TERP for all the $713$ configurations. The generated explanations are robust and TERP identified various regions where different dihedral angles are relevant to predictions. The results are in agreement with previous literature, e.g, the relevance of $\theta$ dihedral angle at the transition state between I and III as reported by Chandler \textit{et al}. \cite{bolhuis2000reaction}. Also, the results intuitively make sense, e.g, we see the VAMPnets state definitions change rapidly near $\phi\approx 0$ and TERP learned that $\phi$ is the most dominant feature in that region.} This shows that VAMPnets worked here for the correct reasons and can be trusted. \sm{In \textbf{Fig.} \ref{fig:vamp_01} (d,e,f) we show TERP results for a specific configuration for which $j=2$ non-zero model resulted in optimal interpretation.} \textbf{Fig.} \ref{fig:vamp_01} (f) clearly shows that ($\theta^{j+1}-\theta^j$) is minimized at $j=2$ and the average of $\theta^{j+1}, \theta{j}$ is taken as the optimal temperature $\theta^o$ for calculating $\zeta^j$ (note the minimum at $j=2$) using \textbf{Eq. }\ref{eq:last}. Additional implementation details are provided in \textbf{Sec. \ref{sec:methods_vamp}}.

Next, we demonstrate the advantages of LDA based similarity measure. \textbf{Fig.} \ref{fig:vamp_01}(e) shows that the LDA projection successfully generated two clusters of datapoints belonging to the in-explanation and not in-explanation classes respectively. These well separated clusters help in computing meaningful and improved distance measure $d$. In \textbf{Figs.} \ref{fig:vamp_01}(f,g) we illustrate the robustness of an LDA implementation against noisy and correlated features and compare results with Euclidean similarity implementation. We generate pure white noise by drawing samples from a normal distribution $\mathcal{N}(0,1)$ and generate correlated data by taking $a_ix_i + b\mathcal{N}(0,1)$ (e.g, $a_i=1.0, b = 0.2$), where $x_i$ are standardized features from the actual data. As shown in \textbf{Figs.} \ref{fig:vamp_01}(f,g), we construct synthetic neighborhoods by combining actual data from the four dihedral angles and adding one pure noise, four pure noise, four correlated features respectively. Since the synthetic features do not contain any information, their addition should not change similarity. Thus we can compare the robustness of a measure by computing average change in similarity per datapoint squared which we call similarity error, $\Delta \Pi \in [0,1]$ as shown in \textbf{Eq.} \ref{eq:sim}.

\begin{equation} 
\begin{aligned} 
\label{eq:sim}
\Delta \Pi = \frac{1}{N}\sum_{i=1}^N (\Pi_i^o - \Pi_i^s)^2
\end{aligned} 
\end{equation}

\begin{figure*}[t!]
    \centering
    \includegraphics[width=\linewidth]{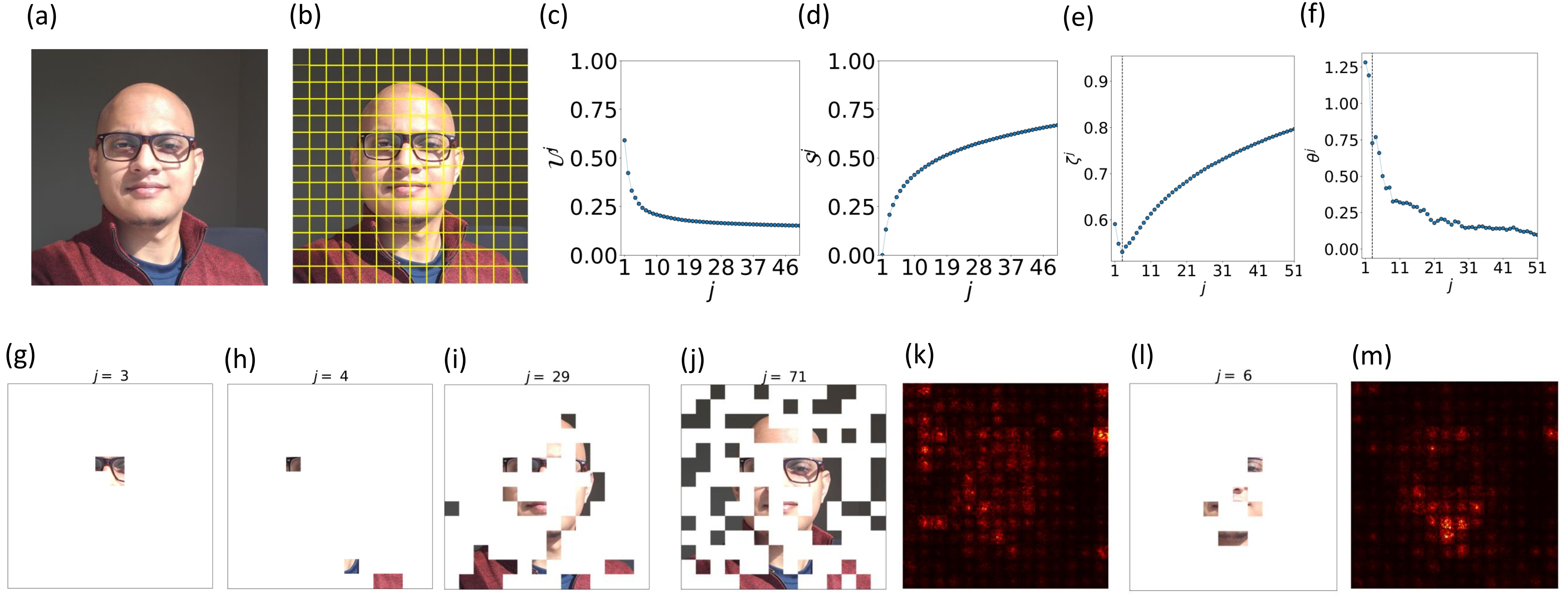}
    \caption{\sm{\textbf{Using TERP to explain and check the reliability of a ViT trained on CelebA dataset.} (a) ViT predicts the presence of 'Eyeglasses' in this image with a probability of $0.998$. (b) Superpixel definitions for the test image following the 16x16 pixel definition of ViT patches. TERP results showcasing (c) $\mathcal{U}^j$, (d) $\mathcal{S}^j$, (e) $\zeta^j$, and (f) $\theta^j$ as a function of $j$, (g) corresponding TERP explanation. We can see the maximal drop in $\theta^j$ happens when going from $j=2$ to $j=3$. By defining the optimal temperature $\theta^o=\frac{\theta^{j=2}+\theta^{j=3}}{2}$ as discussed in \textbf{Sec. } \ref{sec:optimality} a minimum in $\zeta^j$ is observed at $j=3$. Panels (h-j) show sanity checks\cite{adebayo2018sanity} i.e, the result of an AI explanation scheme should be sensitive under model parameter randomization (h,i) and data randomization (j). (k) Saliency map results as baseline explanation for `Eyeglasses' prediction. Red color highlights pixels with high absolute values of the class probability gradient across RGB channels. High gradient at pixels not relevant to `Eyeglasses' shows limitation of saliency map explanation. (i,j) Shows TERP and saliency map explanations for the class `Male'. $\mathcal{U}^j$, $\mathcal{S}^j$, $\zeta^j$, and $\theta^j$ as a function of $j$ plots are provided in the SI.}}
    \label{fig:mobile_01} 
\end{figure*}

Here, the superscripts $o$ and $s$ represent similarities corresponding to the original and synthetic datapoints respectively. From \textbf{Fig.} \ref{fig:vamp_01}(g,h) we can see that LDA based similarity performs significantly better in $100$ independent trials compared to Euclidean similarity. On the other hand, the addition of one pure noise introduces significant similarity error for Euclidean measure. Thus we can conclude that adopting LDA over Euclidean similarity measure will produce significantly improved \smm{explanation}.

\smm{In this section, we demonstrated the applicability of TERP for probing black-box models designed to analyze time-series data coming from MD simulations. In addition to assigning confidence to these models, TERP can be used to extract valuable insights (relevant degrees of freedom) learned by the model. In the future, we expect an increased adoption of TERP-like methods in the domain of AI-enhanced MD simulations for investigating conformational dynamics, nucleation, target-drug interactions, and other relevant molecular phenomena.}

\subsection{Application to Image classification: Vision Transformers (ViTs)}
\label{sec:Image classification}

\sm{Transformers are a type of machine learning model characterized by the presence of self-attention layers and are commonly used in natural language processing (NLP) tasks.\cite{vaswani2017attention} 
The more recently proposed Vision transformers (ViTs)\cite{dosovitskiy2020image} aim to directly apply the transformer architecture to image data, eliminating the need for convolutional layers and have become a popular choice in computer vision. Per construction, ViTs are black-box models and because of their practical usage, it is desirable to employ an explanation scheme to validate their predictions before deploying them.}

\sm{ViTs operate by segmenting input images into smaller patches, treating each patch as a token similar to words in NLP. These patches are then embedded (patch-embeddings) and passed to the transformer layers conducting self-attention and feedforward operations. Such a design allows ViTs to capture long-range spatial dependencies within images and learn meaningful representations. Interestingly, ViTs are known to perform poorly with limited training data but with sufficiently large datasets, ViTs have been shown to outperform convolutional layer-based models. Thus a typical ViT implementation includes two stages: first a large dataset is used to learn meaningful representation and pre-train a transferable model, followed by fine-tuning for specific tasks.}

\sm{In this work, we employ a ViT pre-trained on the ImageNet-21k dataset from the authors \cite{dosovitskiy2020image,steiner2021augreg,rw2019timm} and then fine-tune the model for predicting human facial attributes by training on the publicly available large-scale CelebFaces Attributes (CelebA)\cite{liu2018large} dataset. CelebA is a large collection of $202,599$ human facial images and each image is labeled with 40 different attributes (e.g, `Smiling', `Eyeglasses', `Male', etc.). During training, input images are converted into 16x16 pixel patches resulting in a total of $196$ patches for each CelebA image (224x224 pixel) depicted in \textbf{Fig. } \ref{fig:mobile_01} (b). Other details of the architecture and training procedure are provided in \textbf{Sec. \ref{sec:methods_mobile}}.}

\sm{To explain the ViT prediction `Eyeglasses' (prediction probability of $0.998$) for the image shown in \textbf{Fig.} \ref{fig:mobile_01} (a) using TERP, we first construct human-understandable representative features by dividing the image into $196$ superpixels (collection of pixels) corresponding to the $196$ ViT patches as shown in \textbf{Fig.} \ref{fig:mobile_01} (b). Afterwards, a neighborhood of perturbed images was generated by averaging the RGB color of randomly chosen superpixels following the scheme outlined in \textbf{Sec. }(\ref{sec:neighbor_gen}). \textbf{Fig.} \ref{fig:mobile_01} (c-f) show $\mathcal{U}^j$, $\mathcal{S}^j$, $\zeta^j$, and $\theta^j$ as functions of $j$ after implementing the TERP protocol (\textbf{Algorithm} \ref{alg1}). Thus, TERP explanation enables us to conclude that the ViT prediction of `Eyeglasses' was made for the correct reasons. The optimal TERP explanation shown in \textbf{Fig.} \ref{fig:mobile_01} (g) appears at $j=3$, due to the maximal decrease in $\theta^j$ as $j$ is increased from $2$ to $3$. Using \textbf{Eq. }\ref{eq:temp},\ref{eq:last} $\zeta^j$ is calculated and a minimum occurs at $j=3$.}

\sm{To establish that TERP indeed takes both the input data and the black-box model into account when generating explanations we subject our protocol to the sanity tests developed by Adebayo \textit {et al.} \cite{adebayo2018sanity}. We achieve this by taking the fine-tuned ViT model and randomizing the model parameters in a top-to-bottom cascading fashion following their work and obtain corrupted models. Specifically, we randomize all parameters of ViT blocks $11-9$, and blocks $11-3$ respectively to obtain two corrupt models. TERP explanations for `Eyeglasses' for these two models are shown in \textbf{Fig.} \ref{fig:mobile_01} (h-i). Plots showing $\mathcal{U}^j$, $\mathcal{S}^j$, $\zeta^j$, and $\theta^j$ as functions of $j$ for these models are provided in the SI. Here, the idea is that, due to randomization the explanation will not match the ground truth but a good AI explanation scheme should be sensitive to this randomization test and produce different explanations from the fully trained model. Similarly, we implemented the data randomization test (\textbf{Fig.} \ref{fig:mobile_01} (j)) proposed in the same work, where the labels of the training data are randomized prior to training and a new ViT is trained (training details provided in the SI) using the corrupted data. Again, the results of an AI explanation method should be sensitive to this randomization. From the TERP explanations shown in panels (h-j), we conclude TERP passes both randomization tests.}

\sm{Furthermore, we compute saliency map (\textbf{Sec. }\ref{sec:methods_mobile}) explanations for `Eyeglasses' prediction as baseline. As shown in \textbf{Fig.} \ref{fig:mobile_01} (k), we see the limitations of the saliency explanation e.g, a lot of pixels irrelevant to `Eyeglasses' are detected to have high absolute values of the probability gradient across the RGB channels. This is not surprising since saliency maps are known to detect color changes, object edges, and other high-level features instead of learning a relationship between model inputs and class prediction.\cite{adebayo2018sanity} This analysis serves as a baseline and helps to put TERP explanations into context. Finally, we also generate TERP and saliency map explanations for the label 'Male' as shown in \textbf{Fig.} \ref{fig:mobile_01} (l,m). A more detailed analysis of TERP explanations for this prediction is provided in the SI.} \smm{We would also like to emphasize that in addition to helping to decide whether an image classifier can be trusted or not, 
TERP should be useful for uncovering underlying patterns and biases in the training data.}

\subsection{Application to Text classification: Attention-based Bidirectional Long Short-Term Memory (Att-BLSTM)}
\label{sec:text}

Classification tasks in natural language processing (NLP) involve identifying semantic relations between units appearing at distant locations of in a block of text. This challenging problem is known as relation classification and models based on long short-term memory (LSTM)\cite{yu2019review}, gated recurrent unit (GRU)\cite{chung2014empirical}, transformers\cite{vaswani2017attention} have been very successful in addressing such problems. In this work, we look at the widely used attention-based bidirectional long short-term memory\cite{zhou2016attention} (Att-BLSTM) classifier and apply TERP to explain its predictions.

First, we trained an Att-BLSTM model on Antonio Gulli's (AG's) news corpus\cite{gulli}, which is a large collection of more than $1$ million news articles curated from more than $2,000$ news sources. The labels associated with each news article in the dataset indicate the section of the news source (e.g, World, Sports, Business, or Science and technology) that the news was published in. Afterwards, we employed the trained model and obtained prediction for a story titled ``AI predicts protein structures" published in `Nature’s biggest news stories of 2022'.\cite{nature_news_2022}.

\begin{figure}
    \centering
    \includegraphics[width=\linewidth]{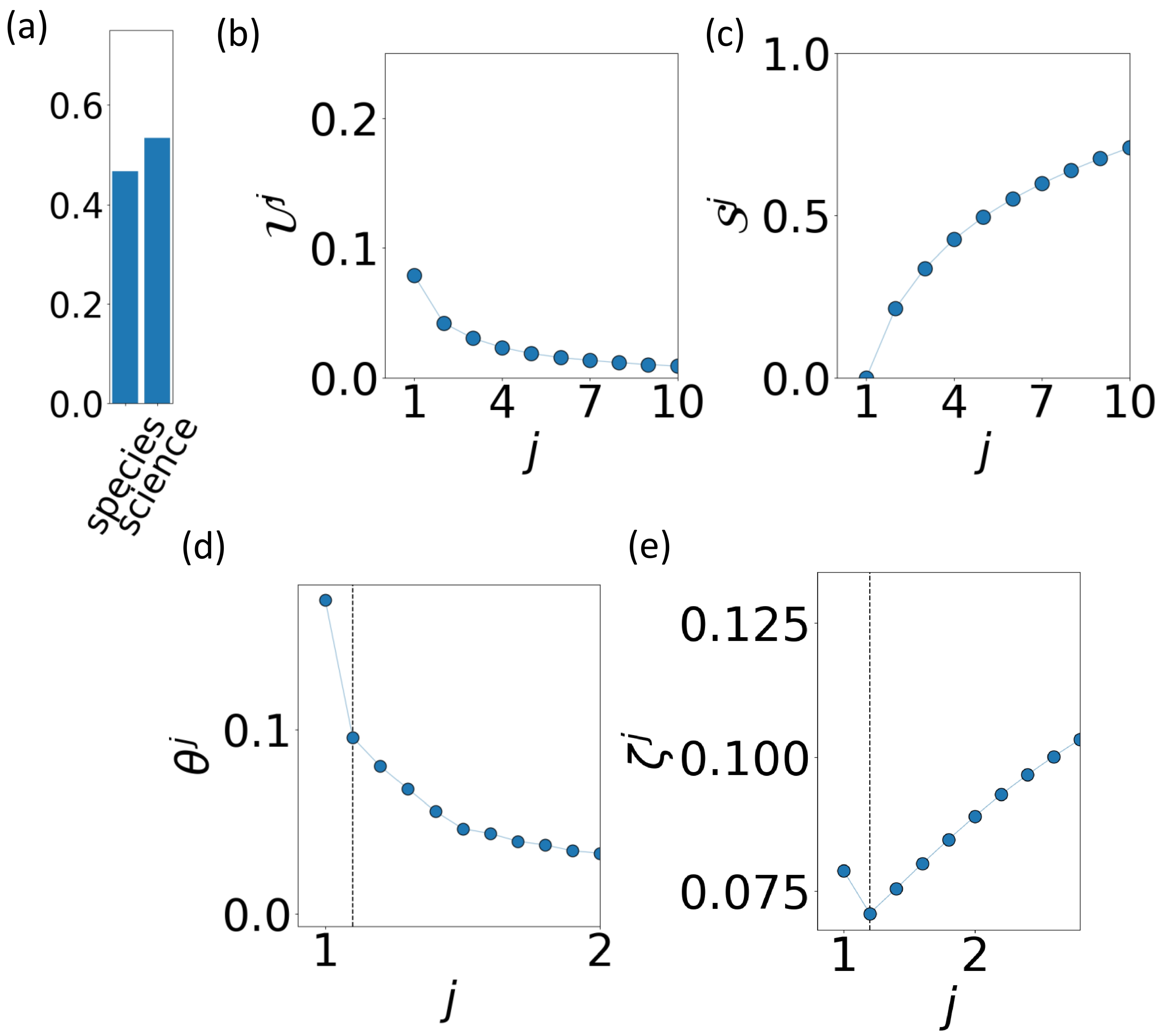}
    \caption{\sm{\textbf{Using TERP to explain and check the reliability of Att-BLSTM model trained on AG's news corpus to predict the news story titled "AI predicts protein structures".} (a) relative feature importance of the two most influential keywords: `science', and `species' as identified by TERP, (b) $\mathcal{U}^j$, (c) $\mathcal{S}^j$, (d) $\theta^j$, (e) $\zeta^j$ vs. $j$ plots showing the optimal explanation appears at $j=2$, due to the maximum drop in $\theta^j$ as $j$ is increased from $1$ to $2$.}}
    \label{fig:text} 
\end{figure}

\sm{To implement TERP for probing a black-box prediction involving text input (sequence of sentences), first the text is passed through a tokenizer (nltk\cite{hardeniya2016natural}) which generates a dictionary of words/phrases contained in that text. These words are the representative features to be used in TERP. Afterwards, a neighborhood of the perturbed text is generated by randomly choosing and removing all instances of different words from the text. TERP processes the neighborhood as numerical values for linear model construction by creating a one-hot-encoded matrix where the columns represent the presence or absence of the different words in a perturbed text.}

As a specific instance, the Att-BLSTM classifier predicted that the story titled ``AI predicts protein structures" is about Science and Technology and we implemented TERP to generate the optimal explanation behind this prediction as shown in \textbf{Fig.} \ref{fig:text}. We see that the most influential keywords are 'species', and 'science' which gives confidence that Att-BLSTM model was able to classify the news story for the correct reasons.

\section{Discussion}
\label{sec:discussion}

The widespread adoption of AI-based black-box models has become a standard practice across various fields due to their ability to be deployed without requiring an in-depth understanding of the underlying processes. However, this advantage also poses challenges regarding trustworthiness and the \smm{explanation} of AI models. In this study, we have introduced a novel framework inspired by thermodynamics to create interpretable representations of complex black-box models. Our objective was to find representations that minimize discrepancies from the true model while remaining highly interpretable to humans using a concept similar to the energy-entropy trade-off. Furthermore, the concept of interpretation entropy introduced in this work has the potential to be useful in more general human interpretability-based model selection from any family of linear models.

We showcased the effectiveness of this approach in various AI applications, including image classification, text analysis, and molecular simulations. While several methods\cite{fisher2019all, lundberg2017unified, sundararajan2017axiomatic, wachter2017counterfactual} have been proposed to address AI interpretability in the past, only a handful, such as Ref. \onlinecite{wellawatte2022model, kikutsuji2022explaining,fleetwood2020molecular}, have been utilized to elucidate molecular simulations. Importantly, our work marks one of the pioneering applications of interpretability techniques in the rapidly evolving field of AI-enhanced molecular dynamics.

Recent applications of our framework (TERP), have been instrumental in uncovering key mechanisms behind crystal nucleation\cite{wang2024local} and hydrophobic ligand dissociation.\cite{beyerle2024thermodynamically} Given the critical role of molecular sciences in uncovering chemical reaction pathways,\cite{yang2017automatic} understanding disease mechanisms,\cite{hollingsworth2018molecular} designing effective drugs,\cite{zhao2015molecular} and numerous other vital areas, it is crucial to ensure accurate analysis, as errors in black-box models can have significant financial and public health implications. TERP should provide practitioners of molecular sciences a way to explain these black-box models on a footing made rigorous through simple yet powerful parallels with the field of thermodynamics.

\section{Methods}
\label{sec:Methods}

\subsection{Neighborhood generation}
\label{sec:neighbor_gen}

We generated neighborhood data around an instance $x=x(x_1,x_2,...x_n)$ by first drawing $n$ numbers from a uniform distribution associated with each feature, $\{t_1,t_2,...,t_n\}\in [0,1]$. The $i^{th}$ feature is perturbed if $t_i\ge 0.5$, otherwise the feature is kept unchanged. Once a feature is chosen for perturbation, the specific scheme for obtaining perturbed values depend on the corresponding data type.

For tabular data if a feature $x$ is continuous, it is updated by $x=x+\epsilon\sigma$ where $\sigma$ is the standard deviation of the feature in the training data and $\epsilon$ is a small noise drawn from a gaussian distribution. For categorial data, feature value $x$ is updated by $x=x'$, where $x'$ is sampled from the training data. For text, an instance is first converted into tokens,\cite{webster1992tokenization} which are considered as features. If a token is chosen for perturbation by following the scheme described above, it is replaced by a new token sampled from training data. For images, superpixels are defined and if chosen for perturbation, are updated by averaging the colors of all the pixels within that particular superpixel.

\subsection{AI-augmented MD method: VAMPnets}
\label{sec:methods_vamp}

The molecular system for alanine dipeptide in vacuum was parametrized using the forcefield CHARMM36m\cite{huang2017charmm36m} and prepared using CHARMM-GUI\cite{lee2016charmm}. A $100$ ns MD simulation of alanine dipeptide in vacuum at $450$ K temperature and $1$ atm pressure was performed using Nose-Hoover thermostat and Parrinello-Rahman barostat\cite{nose1984unified, parrinello1980crystal} in GROMACS.\cite{van2005gromacs}.

A VAMPnets\cite{mardt2018vampnets} deep neural network was constructed from two identical artificial neural network lobes, that take trajectory order parameters (OPs) at time steps $t$ and $t+\tau$ respectively as inputs. The input data was passed through several layers of neurons and finally a VAMP-2 score was calculated by merging results from outputs of both lobes. The neural network model parameters were tuned in successive iterations that maximize the VAMP-2 score. In this way, a markov state model at a specific lagtime $\tau$ can be learnt that describes the slow processes of interest.

In this work, the VAMPnets implementation was performed using the PyEMMA\cite{scherer2015pyemma} (2.5) and Deeptime\cite{hoffmann2021deeptime} 0.4.2 python libraries by constructing the neural network architecture depicted in \textbf{Supporting Fig. 2}. Other training hyperparameters are: $\tau = 0.05 ps, learning\_rate =0.0005, n\_epochs=50$.

\subsection{Image classification: Vision Transformers (ViTs)}
\label{sec:methods_mobile}

Large-scale CelebFaces Attributes (CelebA) Dataset \cite{liu2018large} contains $202,599$ celebrity images, each annotated with $40$ binary attributes. CelebA offers the dataset in two different formats: (a) actual raw images, (b) processed data with aligned facial images. In this work, we employed the latter and divided the dataset into training, validation sets with a ratio of $50:50$. The training data was then used to train a ViT model.

The model was trained until validation metrics (f1 score) did not improve for $5$ consecutive epochs using a learning rate of $0.00001$. The model with the highest validation metric was saved as the trained model.

Training and inference using ViT was implemented using pytorch-lightning 1.5 and python 3.9. The pre-trained ViT model was pulled from the timm python library. For saliency analysis, the absolute values of the gradients of prediction probabilities with respect to input pixels were calculated using the backward() method of pytorch during a backward pass.

\subsection{Text classification: Attention-based Bidirectional Long Short-Term Memory (Att-BLSTM)}
\label{sec:methods:text}

In this work, we employed Python implementation of Att-BLSTM\cite{zhou2016attention} obtained from https://github.com/Renovamen/Text-Classification with pre-trained GloVe word embedding. Att-BLSTM model was trained on Antonio Gulli's (AG's) news corpus\cite{gulli2005anatomy} for $10$ epochs, finally reaching a validation accuracy of $92.0\%$.

\section{Competing Interests}
\label{sec:com_int}
The authors have no competing interests to declare.

\section{Author Contributions}
\label{sec:aut_cont}
P.T. and S.M. designed research; S.M. performed research; S.M. analyzed data; S.M. and P.T. wrote the paper.

\section{Acknowledgments}
This work was supported by the National Science Foundation, grant no. CHE-2044165  (S.M. and P.T.). S.M. was also supported through NCI-UMD Partnership for Integrative Cancer Research. The authors thank Deepthought2, MARCC, and XSEDE (projects CHE180007P and CHE180027P) for the computational resources used in this work.  P.T. is an investigator at the University of Maryland-Institute for Health Computing, which is supported by funding from Montgomery County, Maryland and The University of Maryland Strategic Partnership: MPowering the State, a formal collaboration between the University of Maryland, College Park, and the University of Maryland, Baltimore. P.T. was an Alfred P. Sloan Foundation fellow during the preparation of this manuscript. The authors thank Dr. Eric Beyerle and Dr. Luke Evans for insightful discussions. \newline

\section{Data availability statement}
The data that support the findings of this study are openly available. The AG's news corpus dataset was obtained from \textbf{Ref.} \onlinecite{AGweb}, and CelebA dataset from \textbf{Ref.} \onlinecite{celebaweb} in accordance with the Terms of Service of the respective web resources. Data generation details for the molecular dynamics of alanine dipeptide are provided in \textbf{Sec.} \ref{sec:methods_vamp} and the trajectory is available at \url{figshare.com/articles/dataset/Black-box_models_for_TERP_interpretation/24475003}. Source data is provided with this paper as additional material.

\section{Code availability statement}
Python implementation of TERP is available at \url{github.com/tiwarylab/TERP}, and the three black-box models trained on specific systems studied in this work are available at \url{figshare.com/articles/dataset/Black-box_models_for_TERP_interpretation/24475003}.
\newline \newline

\textbf{References}

	\end{document}